\documentclass[sn-mathphys-num]{sn-jnl}

\usepackage{graphicx}%
\usepackage{multirow}%
\usepackage{amsmath,amssymb,amsfonts}%
\usepackage{amsthm}%
\usepackage{mathrsfs}%
\usepackage[title]{appendix}%
\usepackage{xcolor}%
\usepackage{textcomp}%
\usepackage{manyfoot}%
\usepackage{booktabs}%
\usepackage{algorithm}%
\usepackage{algorithmicx}%
\usepackage{algpseudocode}%
\usepackage{listings}%

\theoremstyle{thmstyleone}%
\theoremstyle{thmstyletwo}%

\theoremstyle{thmstylethree}%

\usepackage{graphicx,amsmath,amssymb,amsbsy,subfigure,hyperref,bbm,times,txfonts,float}
\usepackage{subfigure,hyperref,bbm,times}
\usepackage[T1]{fontenc}
\usepackage{braket}
\usepackage{epsfig} 
\usepackage{color}
\usepackage{graphicx}
\usepackage{dcolumn}
\usepackage{bm}

\providecommand{\openone}{\leavevmode\hbox{\small1\kern-3.8pt\normalsize1}}

\usepackage{soul}

\hypersetup{
   colorlinks=true,
   linkcolor=blue,
}

\begin{document}

\title[Investigating the Impact ...]{Investigating the Impact of Qubit Velocity on Quantum Synchronization Dynamics}


\author[1]{\fnm{Amir Hossein} \sur{Houshmand Almani}}\email{amirhossein.houshmand.almani@gmail.com}

\author*[2]{\fnm{Alireza} \sur{Nourmandipour}}\email{anourmandip@sirjantech.ac.ir}

\author[1]{\fnm{Ali} \sur{Mortezapour}}\email{mortezapour@guilan.ac.ir}

\affil[1]{\orgdiv{Department of Physics}, \orgname{University of Guilan}, \orgaddress{ \city{Rasht}, \postcode{41335-1914}, \state{Guilan}, \country{Iran}}}

\affil*[2]{\orgdiv{Department of Physics}, \orgname{Sirjan University of Technology}, \orgaddress{  \city{Sirjan}, \postcode{7813733385}, \state{Kerman}, \country{Iran}}}


\abstract{We investigate the quantum synchronization dynamics of a moving qubit interacting with a dissipative cavity environment, using the Husimi $Q$-function to analyze its phase space evolution. Unlike conventional synchronization between separate subsystems, we focus on self-synchronization phenomena, where the qubit's phase dynamics exhibit locking to its initial phase distribution. We explore the effects of varying qubit velocity and system detuning across weak and strong coupling regimes. In the weak coupling regime, the system rapidly decoheres with minimal phase preference. In contrast, strong coupling leads to the emergence and persistence of a distinct phase peak, indicating phase locking and enhanced synchronization. These results offer insight into how motion and detuning can regulate coherence and phase stability in open quantum systems. Our approach aligns with recent studies that generalize synchronization concepts to single quantum systems. }




\maketitle

\section{Introduction}\label{sec1}

Synchronization is a fascinating phenomenon in which multiple bodies adjust their motion and rhythms to match through mutual interaction. The study of synchronized dynamical systems dates back to 1673, when Huygens observed the motions of two weakly coupled pendula \cite{75d5327f8ce143f485b91b2a44b6c260}. Synchronization is one of the most widespread behaviors in nature. It can be observed in various systems such as simple coupled pendula, neural oscillations in the human brain \cite{Fell2011TheRO}, epidemic disease spreading in the general population, and power grids \cite{doi:10.1080/00107514.2017.1345844}. Understanding this phenomenon has been a significant achievement of dynamical systems theory and deterministic chaos \cite{BOCCALETTI20021,PhysRevLett.64.821}. From a theoretical point of view, synchronization is a central issue in studying dynamic of chaotic systems and is currently an area of very active research.

Quantum synchronization is typically studied in open quantum systems, as closed systems tend to generate numerous eigenfrequencies with random phase differences, resulting in noisy and indefinite oscillations of observables. Only in large systems, as predicted by the Eigenstate Thermalization Hypothesis \cite{doi:10.1080/00018732.2016.1198134}, do these oscillations dephase rapidly, making observables effectively stationary. Previous studies have mainly focused on the open quantum system, where interactions with the environment result in decoherence and decay within the density operator describing the system state, leading to only a few long-lived oscillating modes. These systems have been studied individually, and various measures of quantum synchronization have been introduced, primarily based on the phase-locking of correlations or Husimi Q-functions \cite{Tindall2019QuantumSE,PhysRevLett.121.063601}.

Open quantum systems, particularly those involving a single qubit, have become a central focus in quantum dynamics research due to their interactions with external environments. These interactions can lead to decoherence and dissipation, fundamentally affecting the qubit's evolution. The study of open quantum systems often distinguishes between Markovian and non-Markovian evolutions \cite{MortezapourBPF17,Mortezapour2017,Nourmandipour2021entanglement,RAFIEE2020126748,PhysRevB.92.064501,PhysRevD.95.025020,mojaveri2022control,taghipour2022dynamics}. In Markovian evolution, the qubit interacts with its environment in a memoryless fashion, where information is lost to the environment irreversibly, and the system's future state depends only on its present state, not on its history. This results in a straightforward, exponential decay of coherence. In contrast, non-Markovian evolution is characterized by memory effects, where information can flow back from the environment to the qubit. This backflow leads to more complex dynamics, where the system's future is influenced by its present and past states, often resulting in non-exponential decay and possible revivals of coherence. These behaviors are driven by factors such as structured environmental spectral densities and strong system-environment couplings, making the study of non-Markovian dynamics crucial for developing robust quantum technologies \cite{breuer2002theory}.

In synchronization, the interplay between quantum coherence and environmental conditions plays a pivotal role. Previous studies have underscored the significant impact of quantum coherence on synchronization dynamics \cite{PhysRevA.105.L020401}, prompting further investigation into the influence of qubit velocity within a leaky cavity. This choice is motivated by recent findings \cite{MortezapourBPF17,Golkar:20} that highlight the beneficial effects of increasing qubit velocity in a leaky cavity environment, enhancing the preservation of quantum coherence despite dissipation and decoherence effects typical of such systems. As qubit velocity rises within this setup, it is anticipated that phase locking and synchronization will be correspondingly enhanced, owing to the strengthened coherence preservation. This research not only illuminates the complex interplay between qubit behavior, environmental factors, and synchronization phenomena but also holds promise for advancing robust quantum technologies, where maintaining coherence and synchronization are critical objectives.

Motivated by prior research and theoretical foundations, we investigate a specific model with the goal of enhancing quantum synchronization. Our model features a qubit, the fundamental unit of quantum information, moving along the z-axis with velocity $v$. The study focuses on exploring how two key parameters, $\beta$ (qubit velocity) and detuning (frequency mismatch), influence synchronization dynamics in both weak and strong coupling regimes. Among the various approaches introduced to measure synchronization \cite{Galve2017,PhysRevResearch.2.043287,e25081116}, we employ the Husimi Q-distribution function to analyze quantum synchronization. The Husimi Q-function is a quasi-probability distribution that characterizes the states of quantum systems in phase space and is widely used in quantum optics and quantum mechanics.

In classical systems, synchronization is often associated with the emergence of limit cycles, where the dynamics of the system exhibit periodic oscillations that lead to phase locking between coupled oscillators \cite{PhysRevResearch.2.043287}. However, in quantum systems such as the one studied in this paper, synchronization can manifest without the presence of limit cycles. In our model, the synchronization-like behavior arises from the interaction of the qubit  with the dissipative cavity environment, which leads to the stabilization of the qubit's phase coherence rather than periodic oscillations. Specifically, as the qubit moves through the cavity and interacts with the cavity modes, changes in the velocity of the qubit, detuning, and coupling strength influence the alignment of the qubit's phase with the environmental modes, resulting in a phase preference. This phase preference, captured by the Husimi Q-function, indicates a form of synchronization, even in the absence of classical limit cycles \cite{ALI2024129436,PhysRevResearch.5.033209,Luo_2025}.

Our findings indicate that in the weak coupling regime, variations in $\beta$ and detuning have negligible effects on synchronization emergence. However, in the strong coupling regime, increasing $\beta$ notably enhances phase locking, thereby promoting synchronization. Conversely, adjusting the detuning mitigates oscillatory effects that can disrupt synchronization, thereby strengthening overall synchronization dynamics in our model. This investigation aims to deepen our understanding of the intricate relationships between parameters critical to optimizing quantum synchronization. Such insights are crucial for advancing applications in quantum computing, communication, and related quantum technologies.

This paper is structured as follows. In Section \ref{model}, we present the model and provide the analytic solutions under the rotating-wave approximation. In Section \ref{sec.QS}, we explore the synchronization features of a moving two-level atom by calculating the Husimi $Q$-function. Section \ref{sec.SM} deals with  how these factors impact the synchronization region. Lastly, in Section \ref{sec.Con}, we offer a brief discussion and summarize our main findings.

\section{The Model}
\label{model}

\begin{figure}[ht]
   \centering
\includegraphics[width=0.8\textwidth]{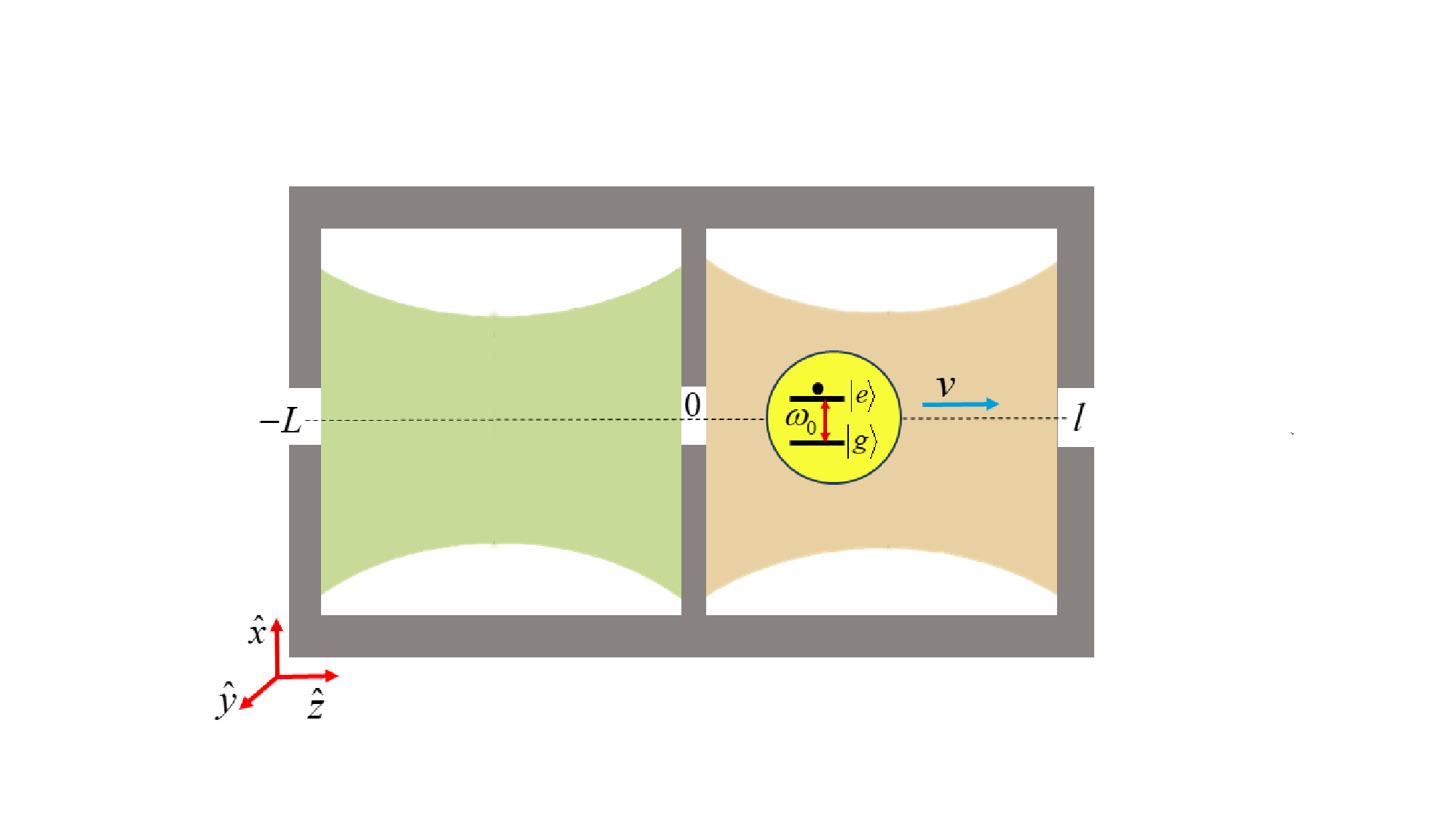}
   \caption{ Pictorial representation of a setup in which a moving qubit is interacting with a dissipative cavity. \label{Fig1}}
  \end{figure}

We have a system consisting of a qubit (a two-level atom with a transition frequency of $\omega_0$), and a structured environment that includes two perfect reflecting mirrors positioned at $z = -L$ and $z = l$, with a partially reflecting mirror at $z = 0$. This arrangement creates two consecutive cavities $(-L, 0)$ and $(0, l)$, as illustrated in Fig. \ref{Fig1}. It is assumed that the qubit interacts exclusively with the second cavity $(0, l)$ and moves along the $z$-axis at a constant velocity $v$. The system's Hamiltonian (with \(\hbar = 1\)) is described as follows:
 \begin{eqnarray}
 \label{eq:model}
   \hat H_{\text{AF}}&=&\hat{H}_{{\text{AF}}}^0+ \hat{H}_{{\text{AF}}}^{\text{Int}},
  \end{eqnarray}
in which
 \begin{subequations}
     \begin{eqnarray} 
   \hat{H}_{{\text{AF}}}^0&=&\sum_{k} \omega_{k} \hat{a}^{\dagger}_{k} \hat{a}_{k}  + \frac{\omega_{\text{0}}}{2}\hat{\sigma}_{z} ,\\
  \hat{H}_{{\text{AF}}}^{\text{Int}}&=&\sum_{k} g_{{k}} f_{k}(z)\hat{a}_{k}\hat{\sigma}_{+}+\text{H.c.}
    \end{eqnarray}
 \end{subequations}

In the given equations, ${{\hat{\sigma }}_{z}}=\left| e \right\rangle \left\langle  e \right|-\left| g \right\rangle \left\langle  g \right|$ represents a Pauli matrix. The symbol ${{\omega }_{k}}$ represents the frequency of the quantized modes of the cavity, and ${{\hat{a}}_{k}}$ ($\hat{a}_{k}^{\dagger}$) are the annihilation (creation) operators of the $k$th cavity mode. The symbol ${{g}_{k}}$ denotes the coupling constant between the qubit and the $k$th mode of the environment. Additionally, ${{\hat{\sigma }}_{+}}=\left| e \right\rangle \left\langle  g \right|$ (${{\hat{\sigma }}_{-}}=\left| g \right\rangle \left\langle  e \right|$) denotes the qubit raising (lowering) operator.
The function $f_k(z)$ describes the position-dependent qubit strength interaction with the cavity modes and is defined as \cite{PhysRevA.70.013414}:

  \begin{equation}
  f_{k}(z)=f_{k}(vt)=\sin[\omega_{k}(\beta t-\Gamma)],
  \label{eq:velshapediss}
  \end{equation}

 In the given expression, $\beta=v/c$ and $\Gamma=L/c$, where $L$ represents the length of the cavity and $c$ is the velocity of light. The presence of the sine term in the above relation is due to the boundary conditions. It is important to note that the qubit's translational motion has been handled classically, i.e., $(z=vt)$. This scenario occurs when the de Broglie wavelength $\lambda_B$ of the qubit is much smaller than the wavelength $\lambda_0$ of the resonant transition \cite{MortezapourBPF17,Mortezapour2017,PhysRevA.101.012331}, meaning $\lambda_B/\lambda_0\ll 1$, implying that $\beta\ll 1$.

Typically, synchronization in classical systems requires an external periodic drive; however, in our model, the motion of the qubit through the optical cavity introduces an effective external influence. This time-dependent interaction with the cavity modes modifies the phase coherence of the qubit, enabling the study of synchronization-like phenomena in this quantum system.

Formally, it is more convenient to work in the interaction picture using:
\begin{equation}\label{int}
 \hat{{\mathfrak F}}_{\text{AF}}= e^{i \hat{H}_{\text{AF}}^0t }\; \hat{H}^{\text{Int}}_{\text{AF}}\;e^{-i \hat{H}_{\text{AF}}^0t }.
  \end{equation}
    After some manipulations, the explicit form of the Hamiltonian in the interaction picture may be obtained as  
      \begin{equation}\label{intpicdiss}
       \hat{{\mathfrak F}}_{\text{AF}}=\sum_{k}g_{{k}} f_{k}(z)\hat{a}_{k}\hat{\sigma}^{+} e^{-i(\omega_{k}-\omega_{\text{0}})t}+\text{H.c.}
        \end{equation}

With the initial state as $\ket{\psi_0}=(C_0\ket{g}+C_1\ket{e})\otimes\ket{0}_R$, the state at any time $t$ becomes:
\begin{equation}
\label{Eq.Psit}
\ket{\psi(t)}=C_0\ket{g}\otimes\ket{0}_R+C_1{\cal E}(t)\ket{e}\otimes\ket{0}_R+\sum_k{\cal G}_k(t)\ket{g}\otimes\ket{1_k}_R
\end{equation}
where, $\ket{0}_R$ is the vacuum state and $\ket{1_k}$ is the state with only one photon in the mode $k$ of the environment.

Using the time-dependent Schr\"{o}dinger equation in the interaction picture, we readily obtain an integro-differential equation for the amplitude ${\cal E}(t)$.
  \begin{equation}
 \dot{\cal E}(t)= -\int_{0}^{t}\! F(t,t'){\cal E}(t')  \, \mathrm{d}t',
 \label{eq:survxidiss}
  \end{equation}
where the kernel $F(t,t')$ has the form
\begin{eqnarray}\label{Correlation functiondiss}
F(t,t')=\sum_{k}|g_k|^2e^{i\delta_k(t-t')}f_k(vt)f_k(vt'),
\end{eqnarray}
in which, $\delta_k=\omega_{\text{0}}-\omega_k$. 

As is seen, ${\cal E}(t)$ depends on the spectral density as well as the shape function of the qubit motion. In the continuum limit for the reservoir spectrum, the
sum over the modes is replaced by the integral
\begin{equation}
\sum_{k}|g_k|^2\rightarrow \int \mathrm{d}\omega J(\omega)
\end{equation}
where $J(\omega)$ is the reservoir spectral density. The imperfect reflectivity of the cavity mirrors results in a Lorentzian spectral density for the cavity \cite{Maniscalco2008}:

\begin{equation}
\label{eq:specden}
J(\omega)=\frac{1}{2\pi}\frac{\gamma\lambda^2}{(\omega_{\text{0}}-\omega-\Delta)^2+\lambda^{2}},
\end{equation}
where $\Delta=\omega_{\text{0}}-\omega_c$ represents the detuning between the center frequency of the cavity modes $\omega_c$ and the transition frequency of the qubit $\omega_{\text{0}}$. The parameter $\gamma$ is associated with the microscopic system-reservoir constant, and $\lambda$ represents the width of the distribution describing the cavity losses.

Using \eqref{eq:specden} in \eqref{Correlation functiondiss}, we arrive at the following expression for $F(t,t')$
  \begin{equation}
  F(t,t')=\frac{\gamma\lambda^2}{2\pi}\int\text{d}\omega\frac{\sin[\omega(\beta t-\Gamma)]\sin[\omega(\beta t'-\Gamma)]}{(\omega_{\text{0}}-\omega-\Delta)^2+\lambda^{2}}e^{-i(\omega-\omega_{\text{0}})(t-t')}.
  \end{equation}
Again, in the continuum limit (i.e., as $\Gamma$ tends to infinity) \cite{park2017protection}, the analytic solution of the above relationship emerges:
  \begin{equation}
  F(t,t')=\frac{\gamma\lambda}{4}e^{-\bar{\lambda}(t-t')}\cosh[\theta(t-t')],
  \end{equation}
  in which $\bar{\lambda}\equiv\lambda-i\Delta$ and $\theta=\beta(\bar{\lambda}+i\omega_{\text{0}})$. Once again $F(t,t')=G(t-t')$, which motivates us to use the Laplace transformation technique. After some straightforward, but long manipulations, we may obtain the analytical solution of \eqref{eq:survxidiss} as follows
  \begin{equation}
  \label{eq:suramplitdiss}
  \begin{aligned}
{\cal E}(t)&=\frac{(q_1+y_+)(q_1+y_-)}{(q_1-q_2)(q_1-q_3)}e^{q_1\gamma t} \\
&+\frac{(q_2+y_+)(q_2+y_-)}{(q_2-q_1)(q_2-q_3)}e^{q_2\gamma t} \\
&+\frac{(q_3+y_+)(q_3+y_-)}{(q_3-q_1)(q_3-q_2)}e^{q_3\gamma t}.
  \end{aligned}
  \end{equation}
  in which the quantities $q_i \ \ (i=1,2,3)$ are now the solutions of the cubic equation
\begin{equation}
\label{eq:cubicdiss}
q^3+2(x_1-ix_3)q^2+(y_+y_-+\frac{x_1}{4})q+x_1(x_1-ix_3)/4=0
\end{equation}
with $y_{\pm}=(1\pm\beta)x_1\pm i\beta x_2-i(1\pm\beta)x_3$ where $x_1=\lambda/\gamma$, $x_2=\omega_{\text{0}}/\gamma$ and $x_3=\Delta/\gamma$.

The explicit form of the qubit's reduced density matrix at any time $t$ is obtained by tracing over the environmental variables. In the computational basis, it is represented as:
\begin{equation}
\hat{\rho}(t)=\begin{pmatrix} \rho_{ee}(0)|{\cal E}(t)|^2 && \rho_{eg}(0){\cal E}(t)\\
\rho_{ge}(0){\cal E}^*(t) && 1-\rho_{ee}(0)|{\cal E}(t)|^2
\end{pmatrix},
\label{QDenMat}
\end{equation}
where $\rho_{ee}(0)=|C_1|^2$, $\rho_{eg}(0)=C_1C_0^*$ and $\rho_{ge}(0)=C_1^*C_0$. 

\section{Quantum Synchronization}
\label{sec.QS}

The Husimi $Q$-function is a quasi-probability distribution in phase space that provides a smoothed, non-negative representation of the  density matrix of a quantum state \cite{scully1997quantum}. It is widely used to visualize and analyze quantum states in terms of phase space variables, offering an intuitive representation that combines classical concepts of position and momentum with quantum mechanics. In the context of quantum synchronization, the $Q$-function plays a crucial role by enabling the visualization of phase coherence and phase locking, which are essential indicators of synchronization. In a system without synchronization, the $Q$-function exhibits a nearly uniform phase distribution, reflecting the absence of a dominant phase preference. However, when synchronization occurs, a distinct peak emerges at a specific phase, signifying phase locking, where the system becomes dynamically attracted to a particular phase. The persistence of this peak over time indicates that the system has reached a stable state in phase space. In our model, the interaction of the moving qubit with the cavity environment induces dynamic changes in the $Q$-function, and the emergence and stability of a phase peak serve as signatures of synchronization. As a quasi-probability distribution, the Husimi $Q$-function bridges classical and quantum descriptions, revealing how synchronization phenomena emerge in quantum systems while simplifying the mathematical complexities of other phase-space representations like the Wigner function. This makes it particularly valuable for analyzing the transition between classical and quantum behavior, as well as for studying the effects of decoherence and noise on synchronization. Through the $Q$-function, one can detect, analyze, and track the synchronization process, gaining deeper insights into the behavior of coupled quantum systems. \cite{PhysRevA.107.022221}. 

For a qubit system the $Q$-function is defined as 
\begin{equation}
Q(\theta,\phi,t)=\frac{1}{2\pi}\bra{\theta,\phi}\rho(t)\ket{\theta,\phi}
\label{QFunction}
\end{equation}
where, $\ket{\theta,\phi}=\cos(\theta/2)\ket{e}+\sin(\theta/2)\exp(i\phi)\ket{g}$ is a spin-coherent states. For the density matrix (\ref{QDenMat}), it is straightforward to derive the explicit form of the Husimi $Q$-function as follows
\begin{equation}
	\begin{aligned}
Q(\theta,\phi,t)&=\frac{1}{2\pi}\left(\cos(\theta) \rho_{ee}(0)|{\cal E}(t)|^2 \right. \\
&+\left.  \sin(\theta) \Re\left( e^{i\phi}\rho_{eg}(0){\cal E}(t)\right) + \sin^2(\theta/2)\right). 
\end{aligned}
\end{equation}

\subsection{Weak Coupling Regime}
\label{sec.weaksyn}

\begin{figure}[h!]
	\centering
	\subfigure[\label{Fig2a} ]{\includegraphics[width=0.35\textwidth]{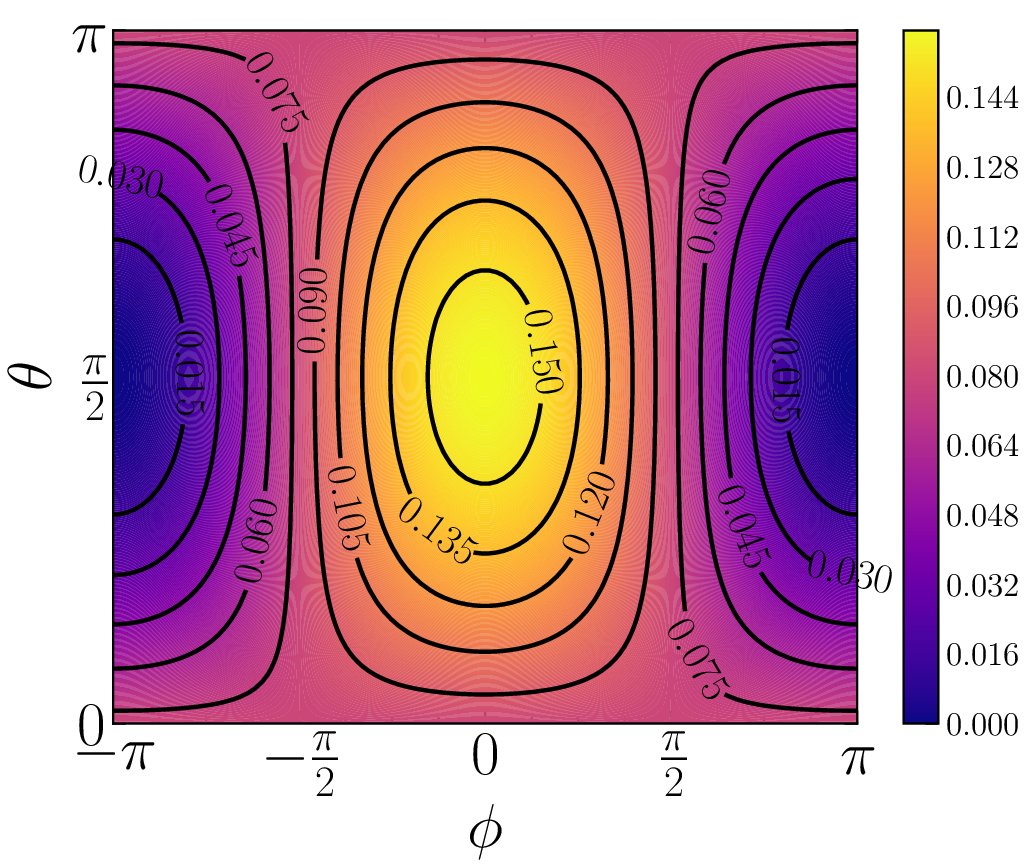}}
	\hspace{0.01\textwidth}
	\subfigure[\label{Fig2b} ]{\includegraphics[width=0.35\textwidth]{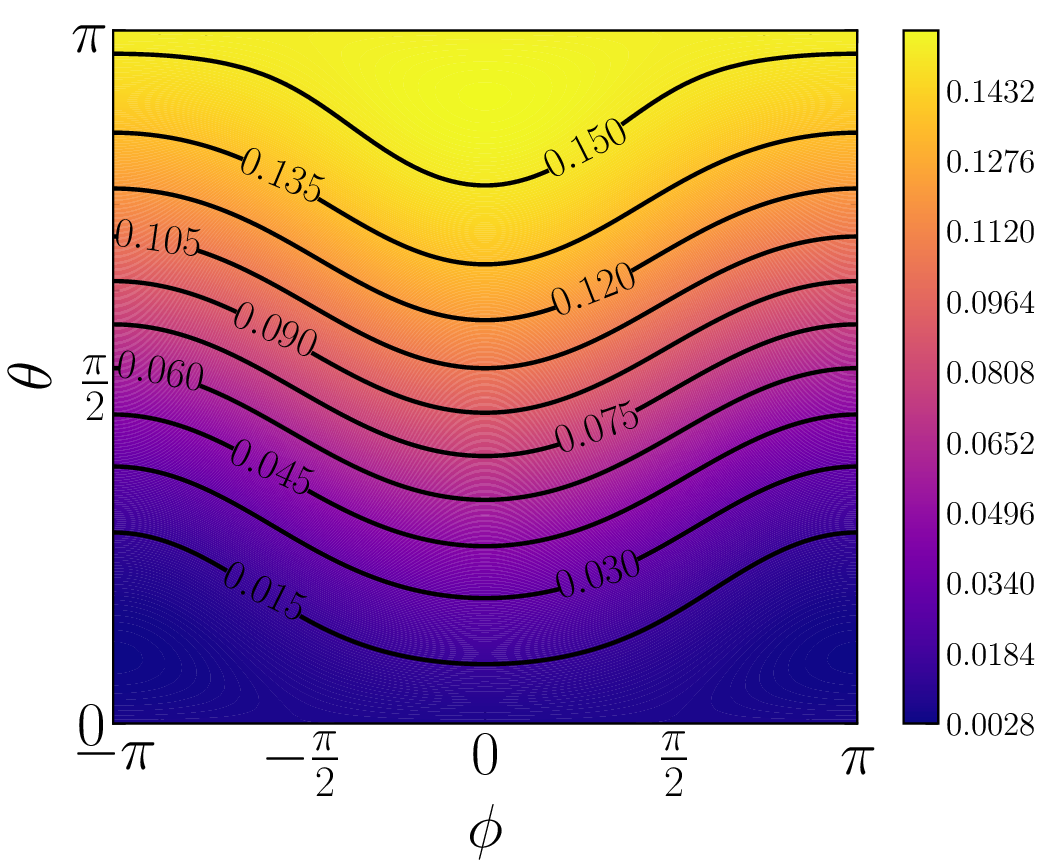}}
	\hspace{0.01\textwidth}
	\subfigure[\label{Fig2c} ]{\includegraphics[width=0.35\textwidth]{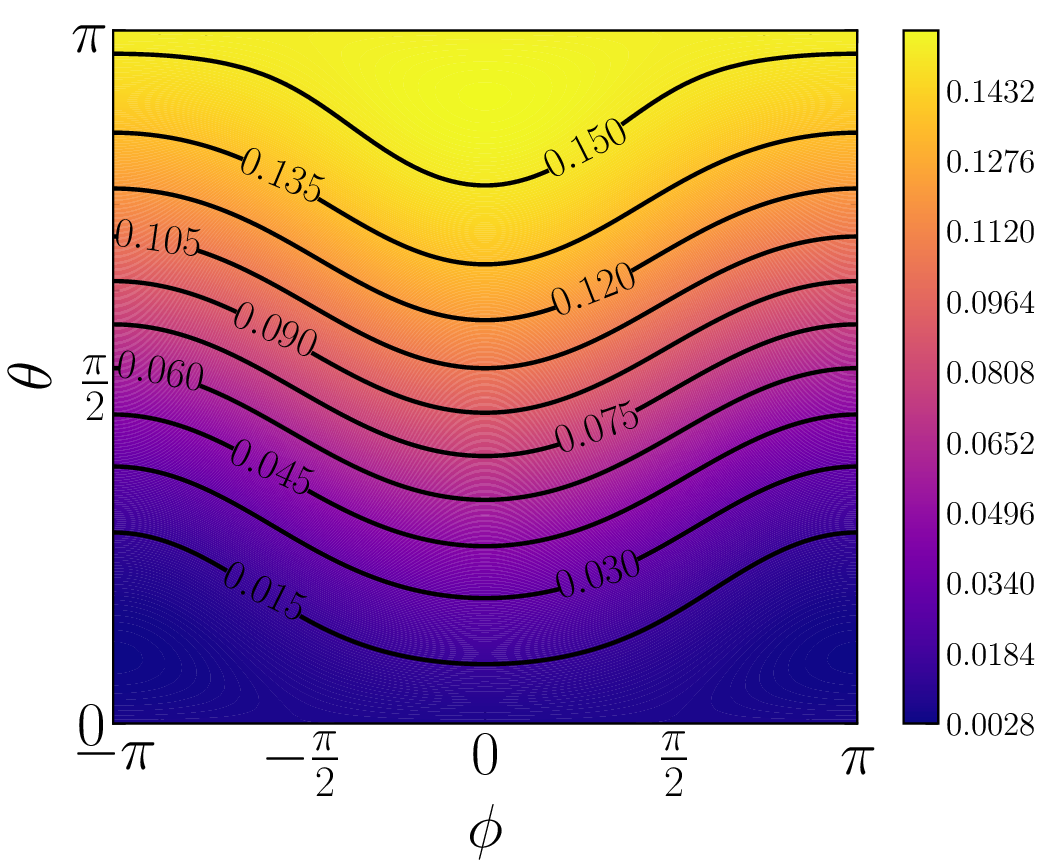}}
	\hspace{0.01\textwidth}
	\subfigure[\label{Fig2d} ]{\includegraphics[width=0.35\textwidth]{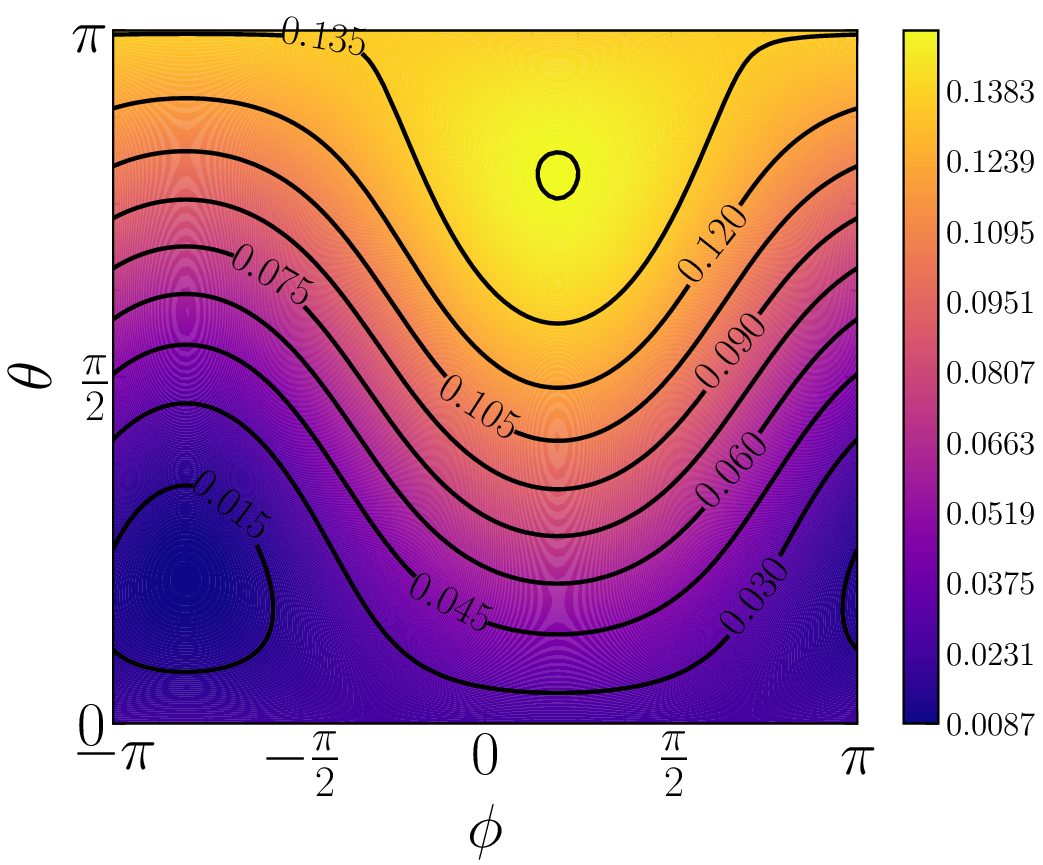}}
	
	\caption{The Husimi function in the weak coupling regime $\lambda = 5\gamma$, for a) the initial time $\gamma t =0$, b) $\gamma t = 5$ with $\beta = 0$ and $\Delta = 0$, c) $\gamma t = 5$ with   $\beta = 0.01\times 10^{-9}$ and $\Delta = 0$ and d) $\gamma t = 5$ with $\beta = 0.01\times 10^{-9}$, $\Delta = 5 \gamma$. The initial state of the system is $\ket{\psi_0}=(\ket{e}+\ket{g})/\sqrt{2}$.} \label{Fig2}
\end{figure}

In Figure \ref{Fig2}, we analyze the behavior of the $Q$-function in the weak coupling regime, specifically when $\lambda = 5\gamma$, for the qubit initial state $\ket{\psi_0}=(\ket{e}+\ket{g})/\sqrt{2}$. At the initial time ($t=0$, Figure \ref{Fig2a}), the $Q$-function exhibits a non-uniform phase distribution, with its maximum value occurring at $\phi=0$. This non-uniformity indicates a phase preference at $\phi=0$, consistent with the findings reported in \cite{PhysRevA.107.022221}. As time progresses, this phase preference diminishes and vanishes by $\gamma t =3$, leading to a uniform $Q$-function distribution along the $\phi$ axis (Figure \ref{Fig2b}). Furthermore, as shown in Figures \ref{Fig2c} and \ref{Fig2d}, introducing a qubit velocity ($\beta = 0.01\times 10^{-9}$) and a non-zero detuning ($\Delta = 5\gamma$) does not preserve the phase preference. In the weak coupling regime, phase information is rapidly lost after a brief evolution period due to the interaction between the system and its environment. Consequently, information lost to the environment does not return to the system to alter its state.

\begin{figure}[h!]
	\centering
	\subfigure[\label{Fig3a} ]{\includegraphics[width=0.35\textwidth]{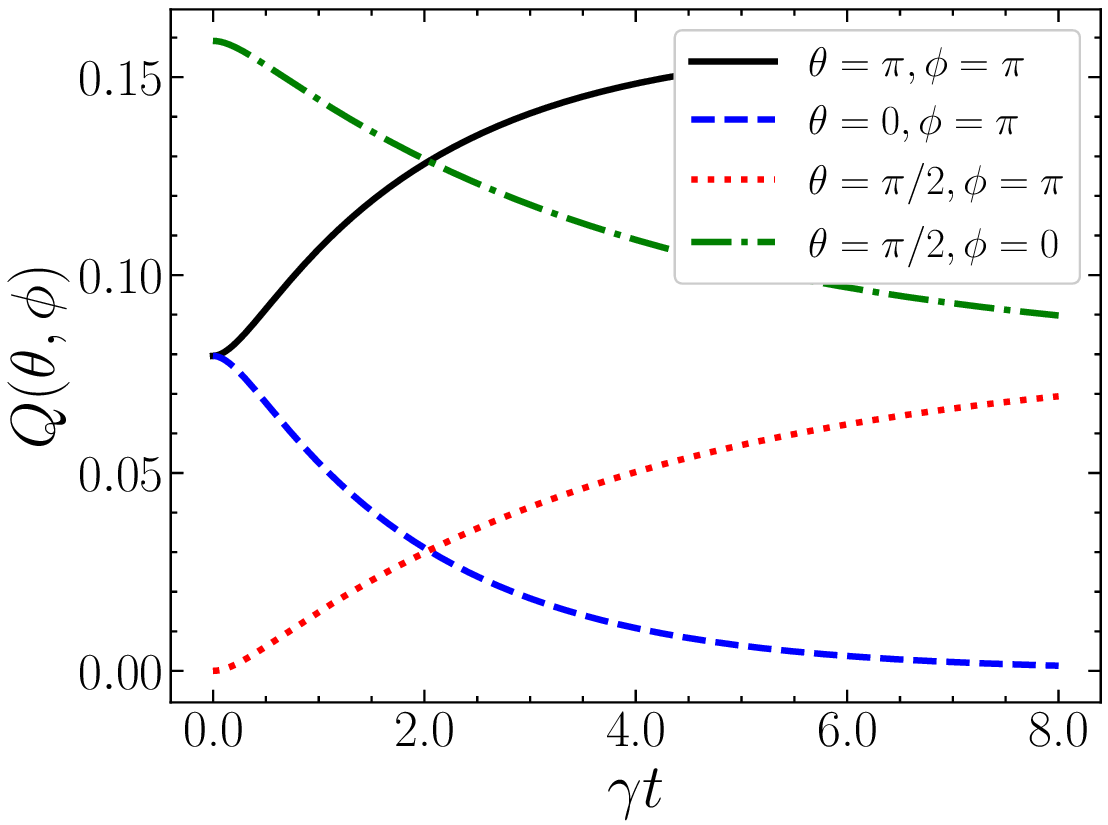}}
	\hspace{0.01\textwidth}
	\subfigure[\label{Fig3b} ]{\includegraphics[width=0.35\textwidth]{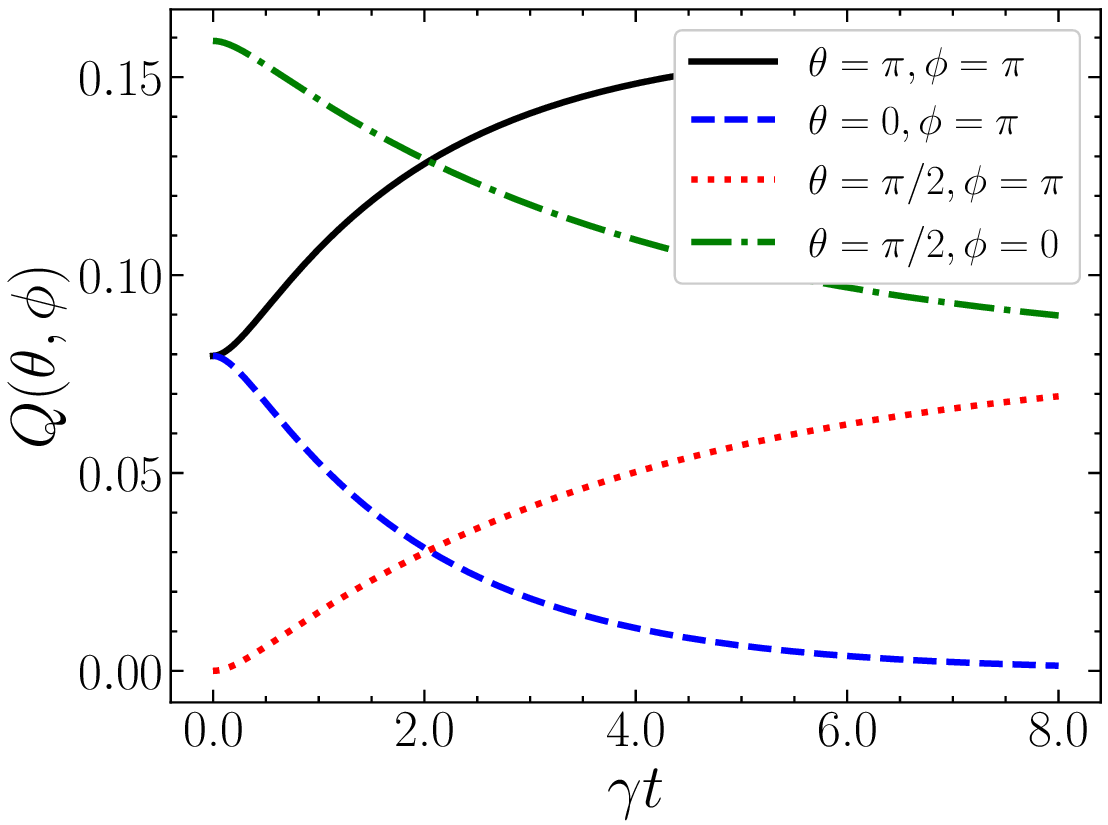}}
	\hspace{0.01\textwidth}
	\subfigure[\label{Fig3c} ]{\includegraphics[width=0.35\textwidth]{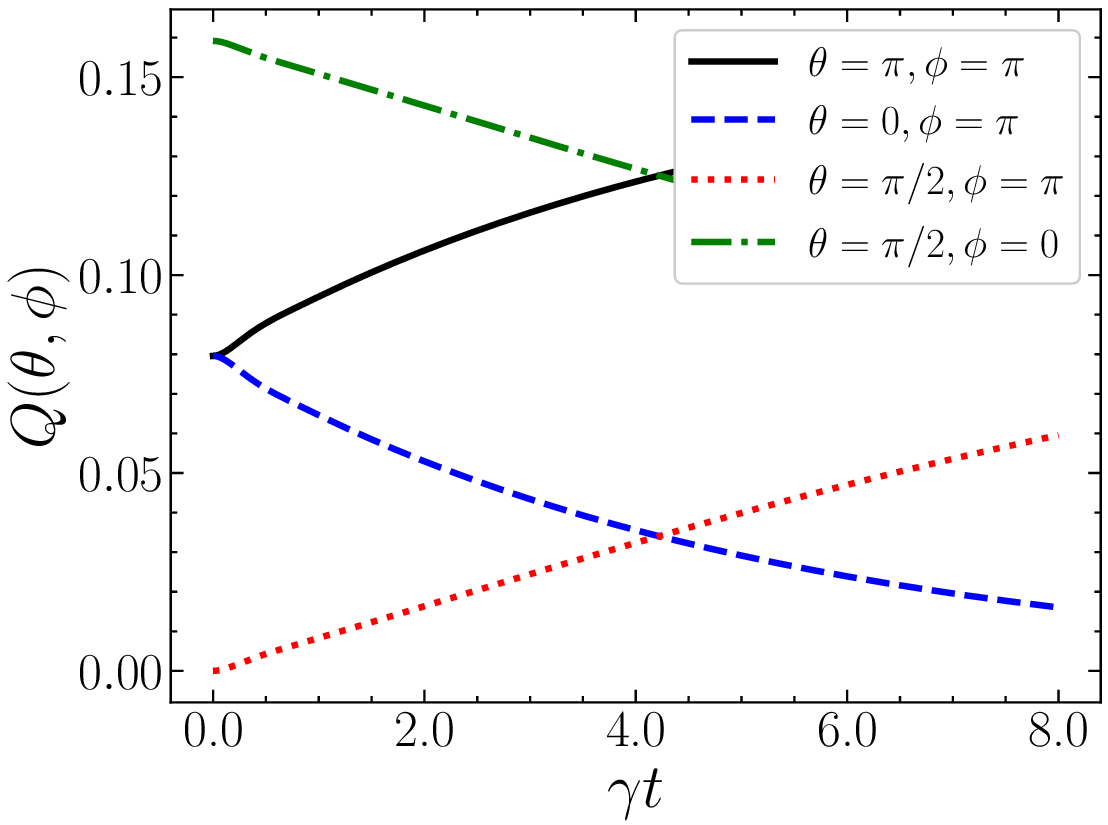}}
	\caption{Time evolution of $Q$ as a function of scaled time $\gamma t$ with $\lambda = 5\gamma$ for a) $\beta = 0$, $\Delta = 0$, b) $\beta = 0.01\times 10^{-9}$, $\Delta = 0$ and c) $\beta = 0.01\times 10^{-9}$, $\Delta = 5\gamma$. The initial state of the system is $\ket{\psi_0}=(\ket{e}+\ket{g})/\sqrt{2}$.} \label{Fig3}
\end{figure}

Figure \ref{Fig3} depicts the dynamics of the $Q$-function in the weak coupling regime ($\lambda = 5\gamma$) for the initial state $\ket{\psi_0} = (\ket{e} + \ket{g})/\sqrt{2}$ across various values of $\theta$ and $\phi$, considering both the presence and absence of qubit velocity and detuning parameters. In these plots, the green dot-dashed line indicates the behavior of the peak of the initial phase distribution for $\theta = \pi/2$ and $\phi = 0$. Conversely, the blue dashed line represents the equilibrium point of the initial phase distribution, corresponding to $\theta = 0$ and $\phi = \pi$. It is evident that, for all parameter values, the $Q$-function uniformly tends towards its distributions, indicating the absence of phase locking or anti-phase locking phenomena.

\subsection{Strong Coupling Regime}

\begin{figure}[h!]
	\centering
	\subfigure[\label{Fig4a} ]{\includegraphics[width=0.31\textwidth]{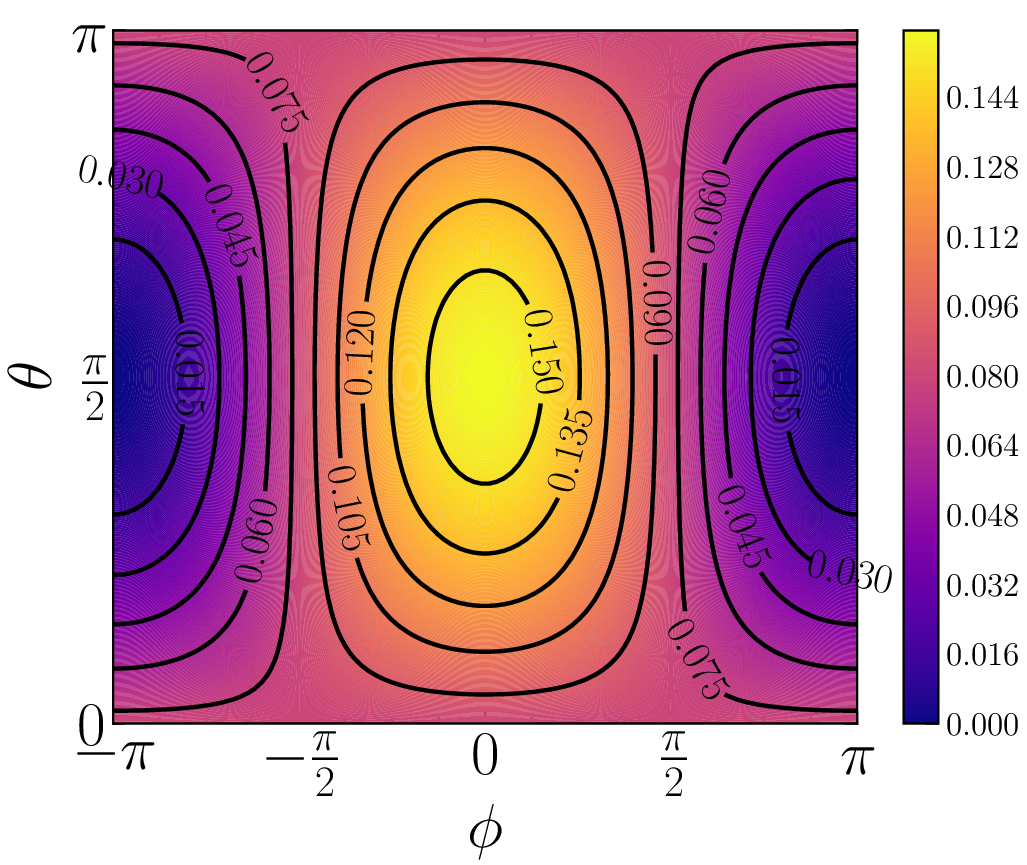}}
	\hspace{0.05\textwidth}
	\subfigure[\label{Fig4b} ]{\includegraphics[width=0.31\textwidth]{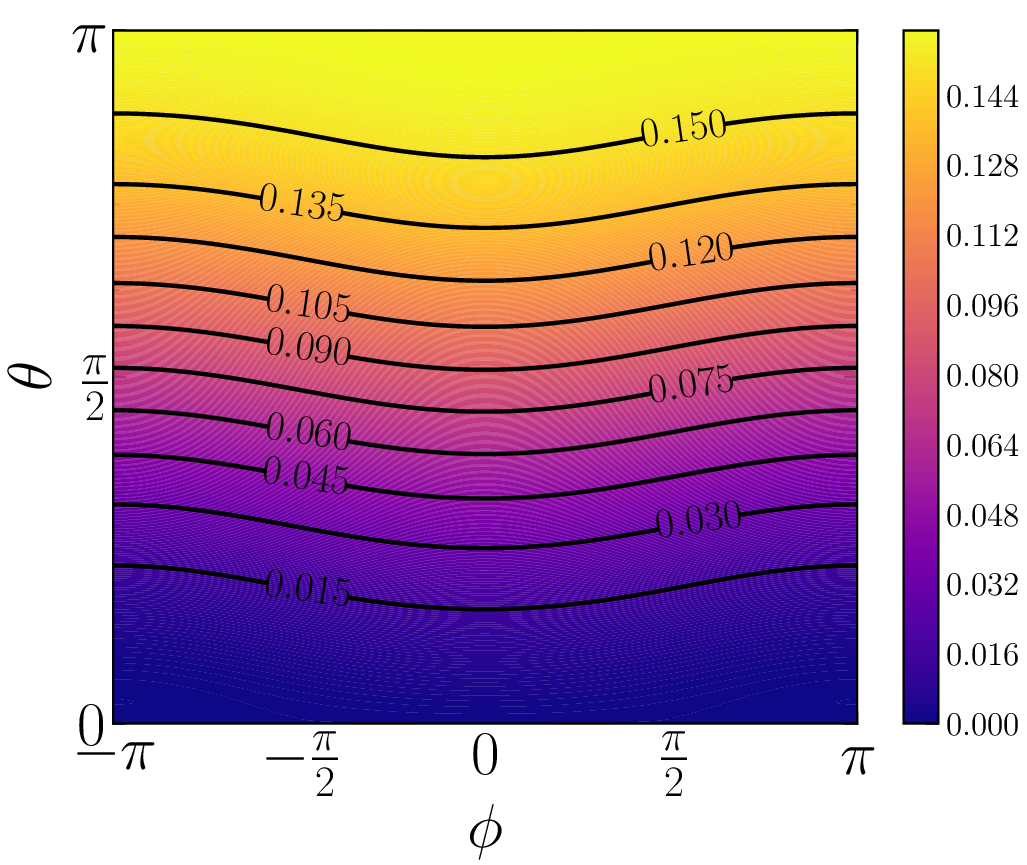}}
	\hspace{0.05\textwidth}
	\subfigure[\label{Fig4c} ]{\includegraphics[width=0.31\textwidth]{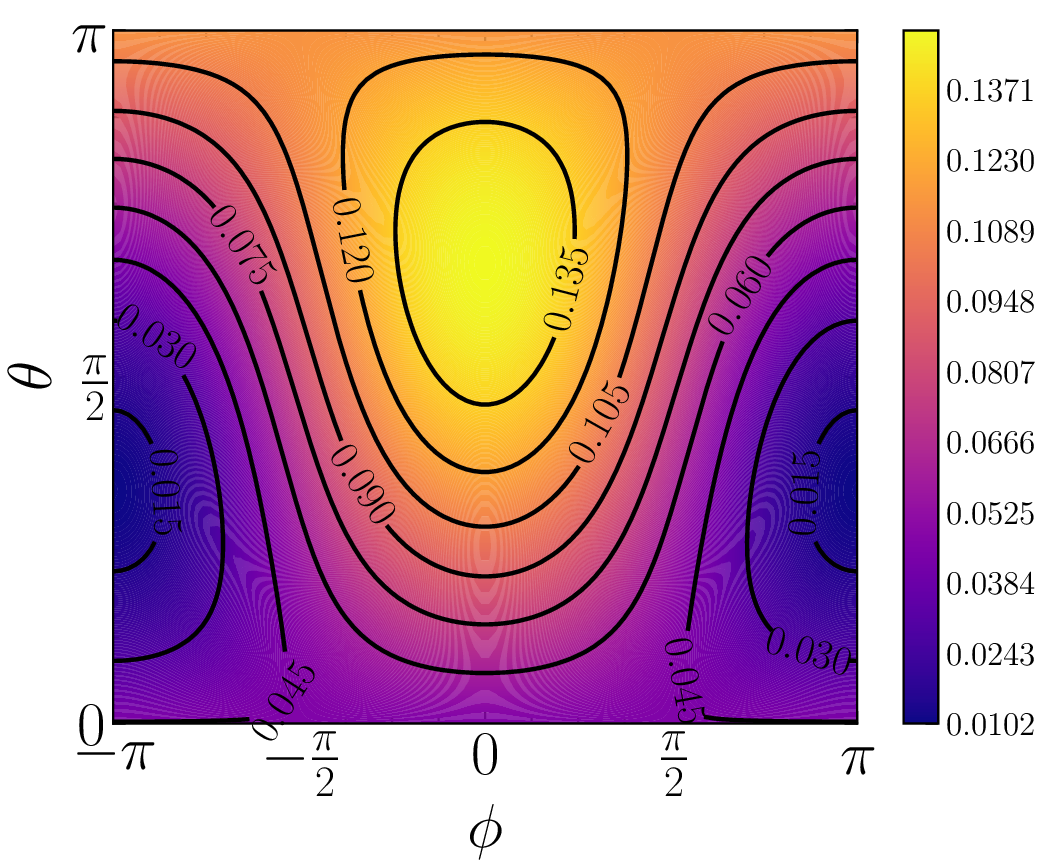}}
	\hspace{0.05\textwidth}
	\subfigure[\label{Fig4d} ]{\includegraphics[width=0.31\textwidth]{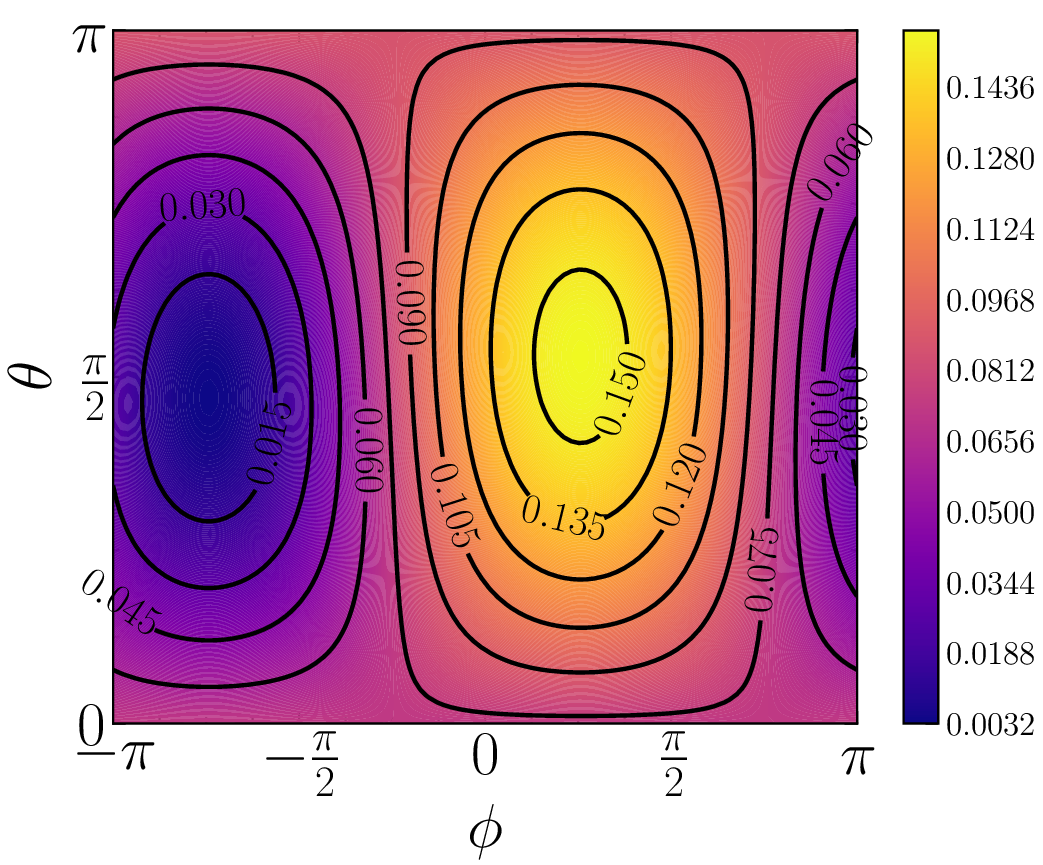}}	
	\caption{The Husimi function in the strong coupling regime $\lambda = 0.01\gamma$, for a) the initial time $\gamma t =0$, b)  $\gamma t = 100$ with $\Delta = 0$ and $\beta = 0.01\times 10^{-9}$, c) $\gamma t = 100$ with $\Delta = 0$ and $\beta = 0.1\times 10^{-9}$ and d) $\gamma t = 100$ with $\beta = 0.01\times 10^{-9}$ and $\Delta = 0.3 \gamma$. The initial state of the system is $\ket{\psi_0}=(\ket{e}+\ket{g})/\sqrt{2}$.}
	\label{Fig4}
\end{figure}

\begin{figure}[h!]
	\centering
	\subfigure[\label{Fig5a} ]{\includegraphics[width=0.31\textwidth]{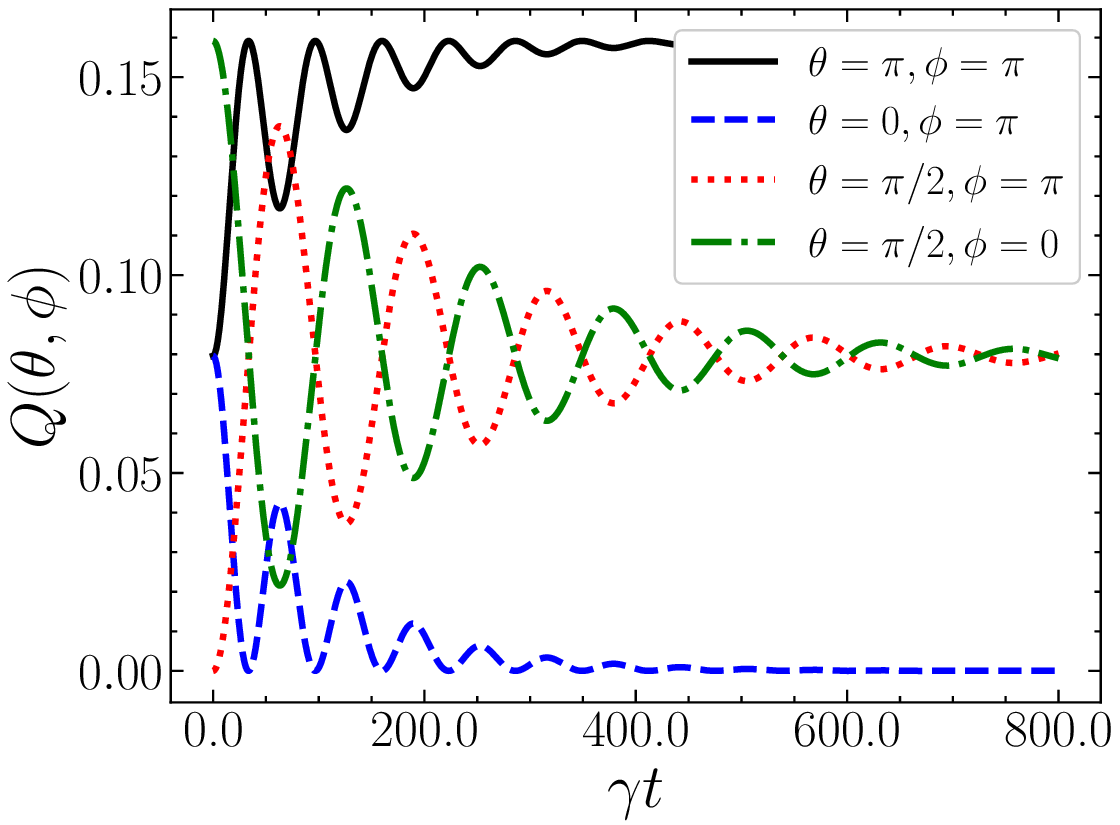}}
	\hspace{0.05\textwidth}
	\subfigure[\label{Fig5b} ]{\includegraphics[width=0.31\textwidth]{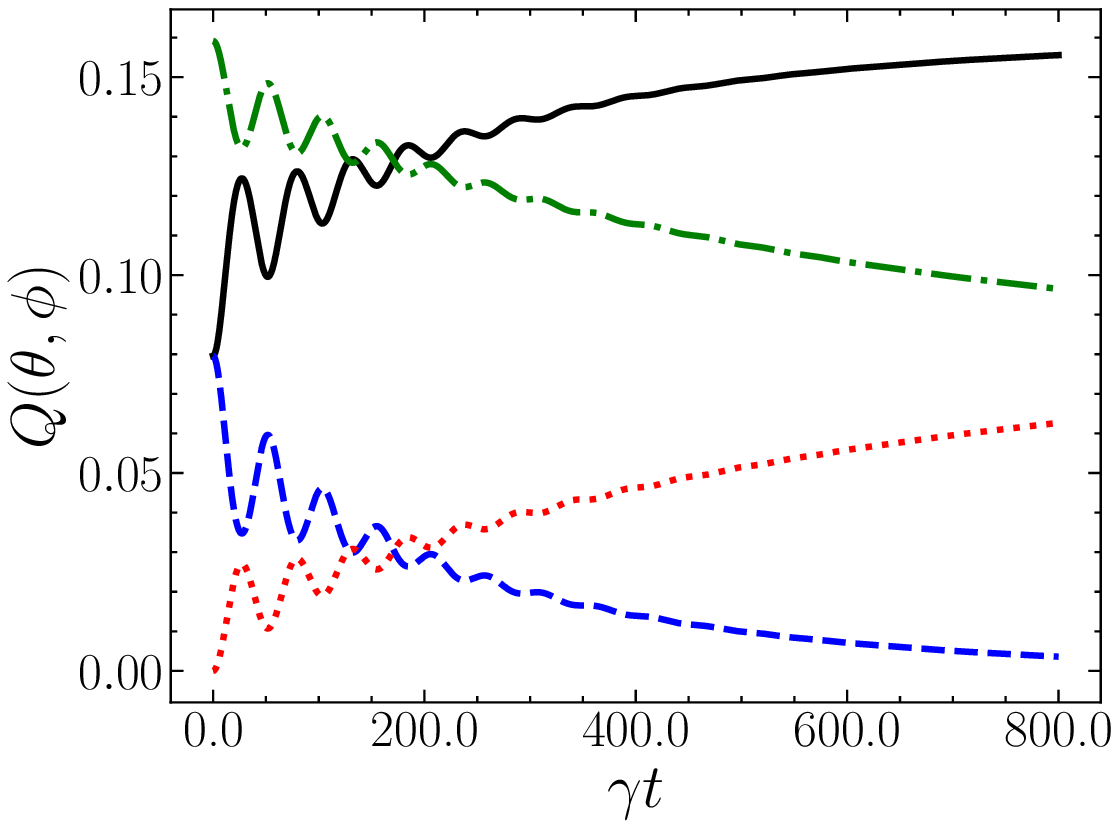}}
	\hspace{0.05\textwidth}
	\subfigure[\label{Fig5c} ]{\includegraphics[width=0.31\textwidth]{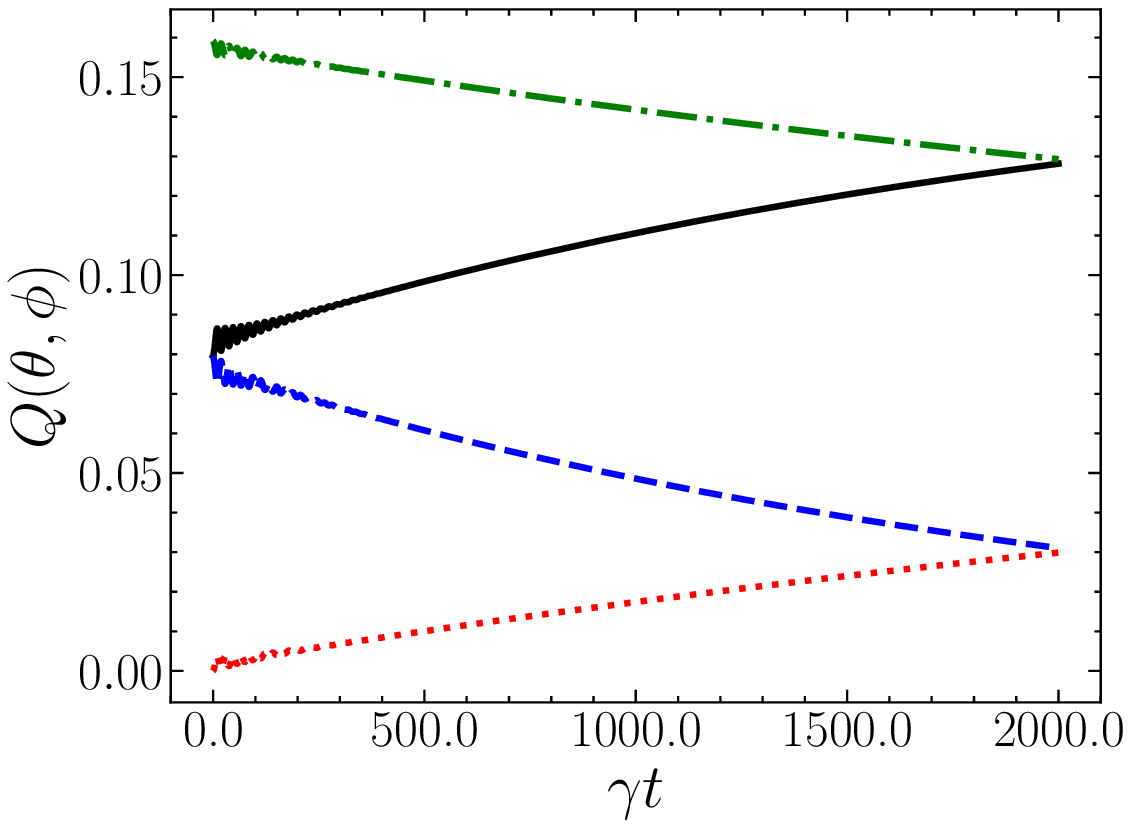}}
	\hspace{0.05\textwidth}
	\subfigure[\label{Fig5d} ]{\includegraphics[width=0.31\textwidth]{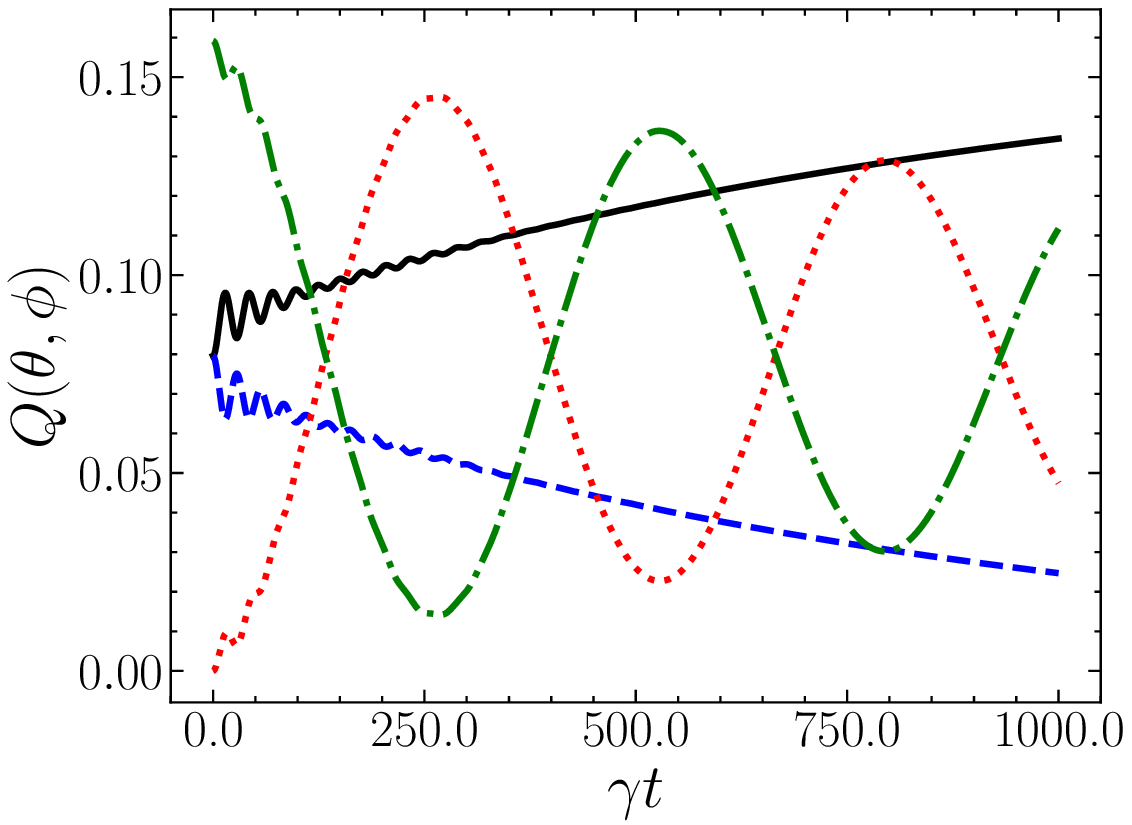}}
	
	\caption{
	Time evolution of $Q$ as a function of scaled time $\gamma t$ with $\lambda = 0.01\gamma$ for a) $\beta = 0$, $\Delta = 0$, b) $\beta = 0.1\times 10^{-9}$, $\Delta = 0$ and c) $\beta = 0.3\times 10^{-9}$, $\Delta = 0$ and d) $\beta = 0.01\times 10^{-9}$, $\Delta = 0.2\gamma$. The initial state of the system is $\ket{\psi_0}=(\ket{e}+\ket{g})/\sqrt{2}$.} \label{Fig5}
\end{figure}

Figure \ref{Fig4} illustrates the $Q$-function in the strong coupling regime with $\lambda = 0.01\gamma$ for the initial state $\ket{\psi_0} = (\ket{e} + \ket{g})/\sqrt{2}$. Similar to the weak coupling regime, the initial phase distribution exhibits a maximum value for $\phi = 0$. As evident from the data presented in Figure \ref{Fig4b}, for small qubit velocities (specifically $\beta = 0.01 \times 10^{-9}$) and zero detuning, this phase preference diminishes and eventually vanishes, resulting in a uniform $Q$-function distribution along the $\phi$ axis. However, higher qubit velocities sustain the initial phase preference for a more extended period, leading to the phenomenon of phase locking (see Figure \ref{Fig4c}). Additionally, non-zero detuning values ($\Delta = 0.3\gamma$) induce dynamical phase locking. Further investigation reveals that the $Q$-function exhibits periodic phase locking and anti-phase locking at specific time intervals. More detailed information is provided in Figure \ref{Fig5c}.

Moreover, this analysis demonstrates that the interplay between qubit velocity and detuning significantly influences the phase dynamics of the system. The study highlights the critical role of these parameters in controlling and manipulating the phase distribution, offering potential applications in quantum information processing and communication.

To gain a more detailed understanding of the behavior of the $Q$-function in the strong coupling regime, we have plotted the dynamics of this parameter as a function of the scaled time $\gamma t$ for the initial state $\ket{\psi_0} = (\ket{e} + \ket{g})/\sqrt{2}$ in Figure \ref{Fig5}. For a qubit at rest ($\beta = 0$) and zero detuning ($\Delta = 0$), the $Q$-function exhibits an oscillatory behavior along with the loss of phase information. This oscillatory behavior arises from the memory depth of the environment, where the strong interaction with the environment forces information to be fed back into the system, thereby altering its state (see Figure \ref{Fig5a}). Furthermore, non-zero qubit velocities significantly impact the phase distribution of the qubit's initial state. For instance, the deterioration of the peak of the initial phase distribution (represented by the green dot-dashed line) is slowed down due to the qubit's velocity (see Figure \ref{Fig5b}). Specifically, for sufficiently large values of $\beta$ ($\beta = 0.3 \times 10^{-9}$), it is evident that the phase difference remains close to zero, resulting in the phase locking phenomenon (see Figure \ref{Fig5c}). The situation changes markedly for non-zero values of the detuning parameter. There is a periodic behavior observed for the peak of the initial phase distribution (green dot-dashed line) oscillating between $\phi = 0$ (corresponding to phase locking) and $\phi = \pi$ (corresponding to anti-phase locking) (see Figure \ref{Fig5d}). These findings underscore the complex interplay between qubit velocity and detuning, and their significant influence on the phase dynamics of the system. The ability to control and manipulate phase distribution through these parameters holds potential applications in quantum information processing and communication. The periodic phase locking and anti-phase locking behavior, in particular, could be exploited for precise timing and synchronization tasks in quantum networks.

\section{Synchronization Measure}
\label{sec.SM}

To quantitatively assess the degree of synchronization in our quantum system, we introduce the synchronization measure. This measure provides a rigorous framework for evaluating how closely the phases of the qubit and its environment align over time. By examining the synchronization measure, we can systematically analyze the impact of varying qubit velocity on the synchronization dynamics. This section will detail the definition and calculation of the synchronization measure, explore its theoretical underpinnings, and present its application in the context of our model. By employing this measure, we aim to gain precise insights into the mechanisms driving phase locking and coherence preservation, which are essential for the effective implementation of quantum synchronization in practical quantum technologies. To this end, we write the synchronization function $S\left( {\phi ,t} \right)$ by integrating the function $Q\left( {\theta ,\phi ,t} \right)$ with respect to $\theta$ as follows:
\begin{equation}
S\left( {\phi ,t} \right) = \int_0^\pi  {d\theta \sin \theta Q\left( {\theta ,\phi ,t} \right)}  - {1 \over {2\pi }},
\label{eq:S}
\end{equation}
which takes the following expression:
\begin{equation}
S\left( {\phi ,t} \right) = {{{\rho _{eg}}\left( t \right){e^{i\phi }} + {\rho _{ge}}\left( t \right){e^{ - i\phi }}} \over 8}.
\end{equation}
The values of $S\left( {\phi ,t} \right)$ can range from $-1$ to $1$, with different values indicating varying levels of synchronization. When $S\left( {\phi ,t} \right)\approx 1$, it signifies strong synchronization, where the phases of the qubit and its environment are closely aligned, exhibiting coherent and synchronized behavior. A value of $S\left( {\phi ,t} \right)\approx 0$ suggests the absence of synchronization, indicating that the phases are uncorrelated and the system exhibits incoherent behavior. On the other hand, $S\left( {\phi ,t} \right)\approx -1$ denotes strong anti-synchronization, where the phases are oppositely aligned, indicating an anti-phase relationship.

\begin{figure}[h!]
	\centering
	\subfigure[\label{Fig6a} ]{\includegraphics[width=0.31\textwidth]{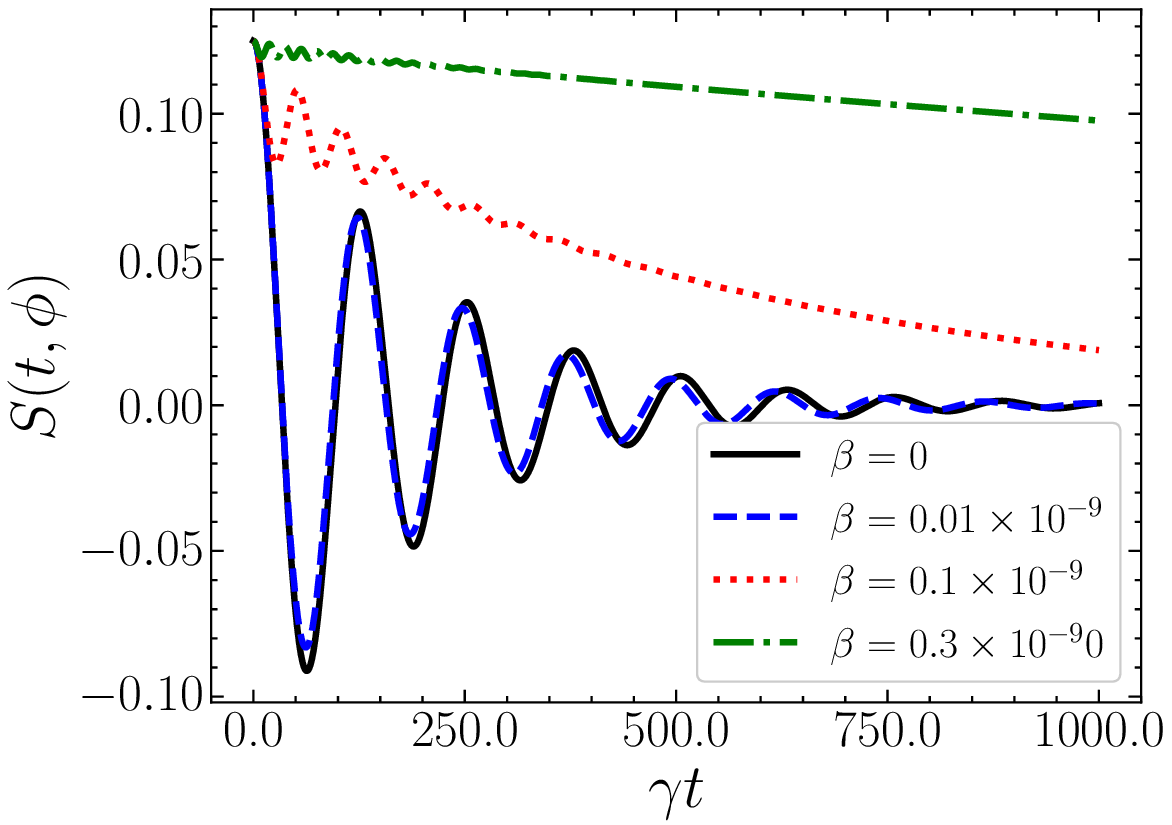}}
	\hspace{0.05\textwidth}
	\subfigure[\label{Fig6b} ]{\includegraphics[width=0.31\textwidth]{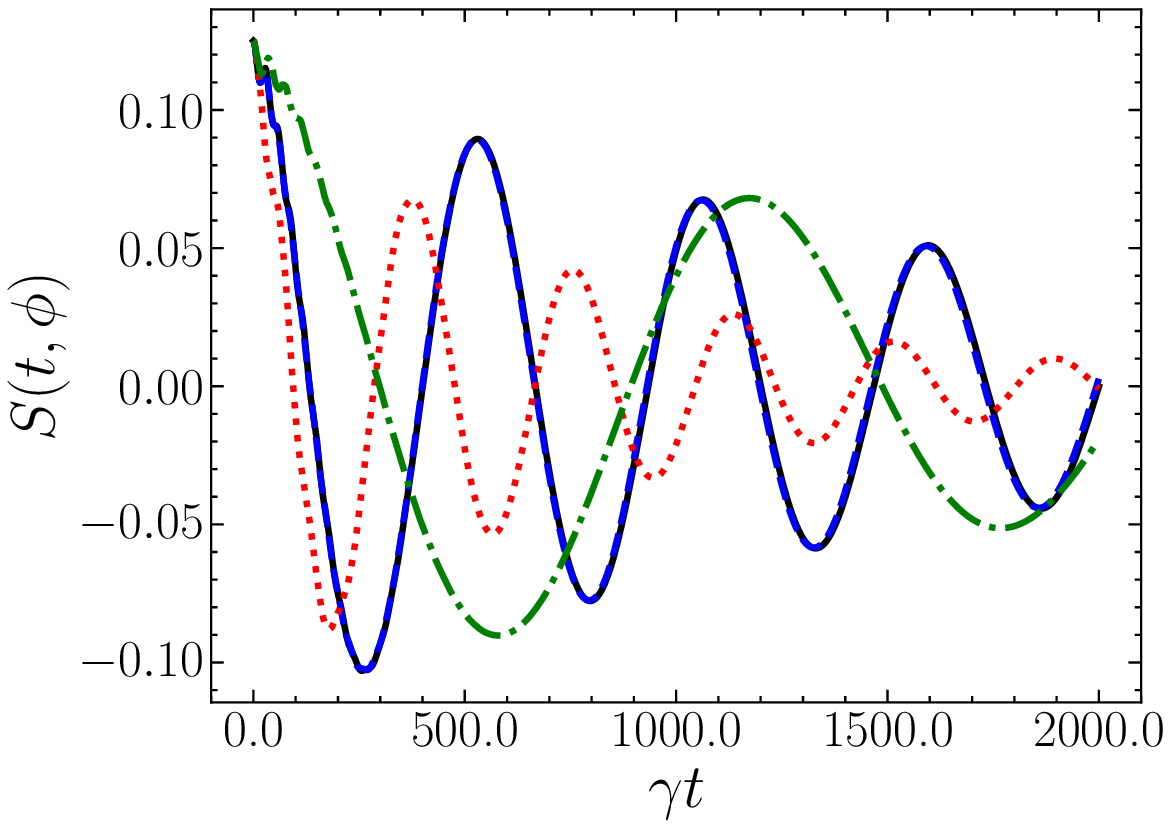}}

   \caption{The synchronization measure $S(t,\phi)$ in the strong coupling regime with $\lambda = 0.01\gamma$ and $\phi = 0$ as a function of $\gamma t$ for (a) $\Delta = 0$ and (b) $\Delta = 0.3 \gamma$. The initial state of the system is $\ket{\psi_0}=(\ket{e}+\ket{g})/\sqrt{2}$.}\label{Fig6}
   \end{figure}

Figure \ref{Fig6} illustrates the dynamics of $S(\phi,t)$ in the strong coupling regime for different values of $\beta$. In the absence of $\beta$ and with $\Delta = 0$, the synchronization measure $S(\phi,t)$ diminishes after a few damped oscillations due to non-Markovian effects (i.e., backflow of information). However, it is observed that for small $\beta$ values (e.g., $\beta = 0.01\times 10^{-9}$) with $\Delta = 0$, oscillations persist, and phase locking does not occur, resulting in a lack of synchronization. Interestingly, increasing $\beta$ can significantly enhance phase locking in the strong coupling regime. The larger the $\beta$, the better the phase locking. On the other hand, when non-zero detuning ($\Delta = 0.3\gamma$) is considered, the intensity of these oscillations is reduced, and phase locking improves as $\beta$ increases. Thus, it can be understood that increasing $\beta$ leads to a significant increase in phase locking and consequently synchronization. Moreover, considering detuning reduces oscillatory effects, further enhancing synchronization.

\section{CONCLUSION}
\label{sec.Con}

This study explores the dynamics of a moving qubit interacting with a dissipative cavity, focusing on the phenomenon of quantum synchronization. Using the Husimi $Q$-function, we visualize the phase space dynamics, offering insights into how the system behaves under different coupling strengths and qubit velocities.

In the weak coupling regime, the $Q$-function analysis showed a quick loss of phase coherence, with no significant phase locking observed. This is due to the weak interaction with the environment, which does not support the retention or feedback of phase information to the system. In contrast, the strong coupling regime shows more complex behaviors, such as oscillatory dynamics and the possibility of phase locking. This is especially apparent at higher qubit velocities or in the presence of detuning. These findings emphasize the significance of environmental memory effects and the strength of the system-environment interaction in controlling synchronization phenomena in quantum systems.

The findings mentioned have important implications for the design and control of quantum systems. This suggests that by precisely adjusting parameters such as the qubit's velocity and detuning, it is possible to manage the coherence and synchronization properties of quantum states. Such control is especially relevant in quantum computing, where maintaining coherent states is crucial, and in quantum communication systems, where synchronized quantum states can improve the accuracy and security of information transfer.

Furthermore, the ability to control synchronization through environmental parameters opens new avenues for developing quantum sensors and metrology tools \cite{vaidya2024quantumsynchronizationdissipativequantum}, where precise measurements rely on stable quantum states. By understanding the conditions that lead to optimal synchronization, we can design better systems for these applications.

In summary, this study not only advances our theoretical understanding of open quantum systems and quantum synchronization but also suggests practical applications in quantum computing, quantum communication, and quantum metrology. Future work could explore the impact of more complex environmental models, including those with non-Markovian characteristics, to further understand the intricate dynamics of quantum synchronization and decoherence. Additionally, investigating the role of different initial states and external controls could provide deeper insights into optimizing and stabilizing quantum systems for practical applications.

\medskip




\bigskip

\bibliography{mybibb}

\end{document}